\documentclass[aip, floatfix, reprint, %superscriptaddress, 
showpacs]{revtex4-1}

\usepackage{graphicx}
\begin{document}




%\preprint{}

\title{Plasmon ruler with gold nanorod dimers: utilizing the second-order resonance} %Title of paper


% , \thanks, \homepage, \altaffiliation all apply to the current author.






\author{Anton T. Le}
\affiliation{Faculty of Physics, Lomonosov Moscow State University, Moscow 119991, Russia}
\author{Maxim R. Shcherbakov}
\affiliation{Faculty of Physics, Lomonosov Moscow State University, Moscow 119991, Russia}
\author{Natalia Dubrovina}
\affiliation{Univ. Paris-Sud, Institut d'Electronique Fondamentale, UMR 8622, 91405 Orsay Cedex, France}
\author{Anatole Lupu}
\affiliation{Univ. Paris-Sud, Institut d'Electronique Fondamentale, UMR 8622, 91405 Orsay Cedex, France}
\author{Andrey A. Fedyanin}
\email{fedyanin@nanolab.phys.msu.ru}
\affiliation{Faculty of Physics, Lomonosov Moscow State University, Moscow 119991, Russia}

%\thanks{}





\date{\today}

\begin{abstract}
The idea of utilizing the second-order plasmon resonance of the gold nanorod $\pi$-dimers for plasmon rulers is introduced. We report on a qualitatively different dependence of the plasmon resonance shift on the interparticle distance for the first- and second-order longitudinal modes, extending the working range of plasmon rulers up to the distance values of 400\,nm.
\end{abstract}

\pacs{36.40.Gk, 73.20.Mf, 78.67.Bf, 78.20.Ci}% insert suggested PACS numbers in braces on next line

\maketitle %\maketitle must follow title, authors, abstract and \pacs

Optical properties of gold nanoparticles have contributed to many areas of science and technology, such as drug delivery \cite{Brown}, cell imaging \cite{Perrault}, photothermal therapy \cite{Huang} and others. In particular, a possibility of measuring nanoscale length utilizing pairs of gold nanoparticles---i.e., plasmonic dimers---was demonstrated \cite{Reinhard,JainEquation} producing the idea of so-called plasmon ruler. Operating principles of plasmon rulers are based on the fact that the spectral position of the surface plasmon resonance (SPR) of a plasmonic dimer strongly depends on the distance between the particles forming the dimer \cite{Rechberger}. This phenomenon makes it possible, e.g., to measure length of a sub-100-nm macromolecule with gold nanoparticles bound to its ends  by measuring the SPR position and comparing it to one of the uncoupled nanoparticles\cite{Reinhard}. Tailorable nature of SPR allows flexible tuning of plasmonic dimer optical properties by changing the shape, mutual arrangement, or polarization of the incoming light and, hence, the type of the resonance \cite{ChiayangTsai,Tabor,ShuchunYang}. In plasmon rulers, however, the maximum measurable distance is still limited to roughly 100\,nm.

Nanorods are one of the frequently used shapes to analyze the physics behind the near-field coupling of plasmonic nanoparticles \cite{Funston}. One distinguishes between two types of nanorod dimers---the so-called $\pi$-dimer and $\sigma$-dimer, referring to the analogy from the orientation of coupled atom $p$-orbitals. Providing subwavelength field localization in the gap between the inline arranged nanoparticles, the $\sigma$-dimer is used in nanoantenna research \cite{ZhengtongLiu} and applications, such as improved surface-enhanced Raman scattering \cite{Alexander}. The $\pi$-dimer, on the other hand, is a system conventionally considered for observation of optical magnetism happening when the free-electron currents are out of phase in the nanorods forming the dimer \cite{Podolsky}.
Finally, for nanorods long enough it is possible to excite several longitudinal SPR modes \cite{Payne}. Nevertheless considered theoretically \cite{JainNR,Willingham},  the coupling properties of the higher-order resonances were neither provided experimentally so far, nor were they considered as a basis for a qualitatively different plasmon ruler-type system.

In this contribution, we obtain the dependence of the SPR position $\lambda_0$ on the distance between the rods $d$ for a set of $\pi$-dimers by means of microspectroscopy of gold nanorod samples and corresponding numerical calculations. The first- and second-order longitudinal dipolar plasmon resonances are considered. It is shown that one can use the   $\lambda_0(d)$ dependence for the second-order resonance to form a plasmon ruler with working distance range of up to 400\,nm. On the contrary, far-field interference effects hinder the possibility of using this system as a plasmon ruler for the same working range if one considers the first-order resonance.

\begin{figure}
\includegraphics[width=\columnwidth]{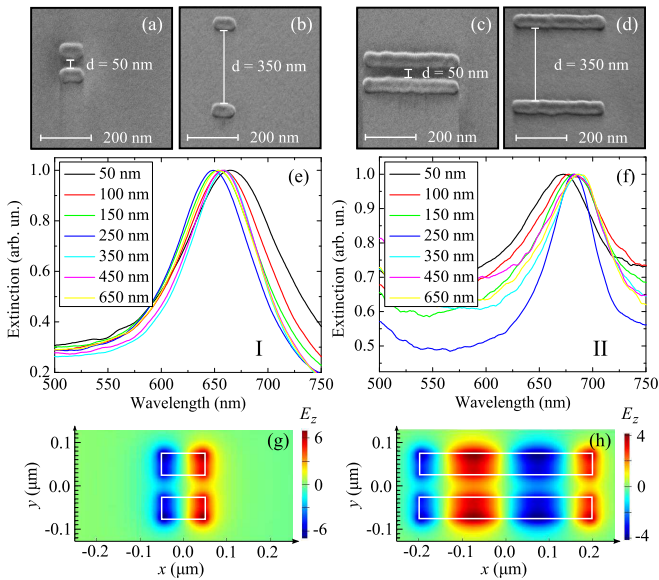}
\caption{(a-d) SEM micrographs of gold nanorod $\pi$-dimers with a nanorod length of $a=100$\,nm or $a=400$\,nm, and an interparticle distance of $d=50$\,nm or $350$\,nm. (e-f) Normalized experimental extinction spectra of nanoparticle dimer sets with a nanorod length of $100$\,nm and $400$\,nm corresponding to the first- (I) and second-order (II) resonances, respectively. (g-h) Calculated $E_z$ distribution nearby the nanorod dimers for the first- and second-order longitudinal resonances, respectively, in the cross-section close to the surface of the dimer.}
\label{fig1}
\end{figure}

The aim of the present study is to perform a comparative study of a dimer plasmon rule operating at the first- or the second-order dipolar resonance. We designed and fabricated two sets of plasmon ruler structures with significantly different dimensions of the resonant elements but providing very close resonance frequencies when operating accordingly to the design at the first- or second-order resonance. Gold nanoparticle dimer samples were fabricated using electron beam lithography with positive electron resist. All the measured nanostructures were located on the same 0.5\,mm-thick fused silica  substrate.  The nanorods were 50\,nm in height, 50\,nm in width; their length $a$ and the distance $d$ between the edges of nanorods within the dimer  varied from sample to sample and constituted a table of parameters with $a=50,$ 100, 150, 200, 300, 400, and 500\,nm, and $d= 50,$ 100, 150, 250, 350, 450, and 650\,nm. Examples of SEM images are shown in Figs.\,\ref{fig1}(a-d) for different combinations of $a=100$\,nm, $a=400$\,nm, $d=50$\,nm, and $d=350$\,nm. Each nanostructured area had lateral dimensions of $100\times100$\,$\mu$m$^2$ and contained approximately 10$^4$ periodically arranged nanorod dimers. The separation distance between adjacent dimers was of approximately 1.5\,$\mu$m; the density of each dimer array was chosen intentionally sparse in order to minimize the interaction effects between the nearest neighbors.

Extinction spectra of gold nanorod dimers with different nanorod length and interparticle distance values were obtained by means of transmission microspectroscopy technique. Transmission spectra were carried out under normal incidence using the white-light microspecrtoscopy setup with a spectral range from 400 to 1400\,nm, a spectral accuracy of 0.8\,nm, a focal spot of 50\,$\mu$m, and a numerical aperture of the focusing system of 0.04. The input polarization state was controlled by a broadband Glan-Taylor prism polarizer to be parallel to the nanorods.

The transmission spectra of all experimentally studied nanorod dimers were also numerically simulated using the finite-difference time-domain (FDTD) technique. Simulations were performed using FDTD Solutions software from Lumerical Solutions, Inc. Gold nanorods were modeled as cuboids with the width and height fixed at 50 and 30\,nm, respectively. The minimal number of grid points per wavelength was set to 22,  and an additional 5\,nm-mesh was implemented in the area of nanorods. The material dielectric constants were taken from Ref.\cite{Palik}. The refractive index of the surrounding medium was set to be 1.0. The incident electromagnetic field was polarized parallel to the nanorods. Unless otherwise noted, the perfectly matched layer (PML) boundary conditions are meant for each side of the integration volume.

% Two different polarization directions of the incident light were chosen, i.e., one parallel to the interparticle axis and the other perpendicular to the axis. Under parallel polarization the plasmon resonance red-shifts as the interparticle distance is reduced. Conversely, there is a blue-shift with decreasing the distance for orthogonal polarization.

%%%In this section the following methods are described: EBL, microspectroscopy, FDTD, DDA.%%%

%\section{results}

The experimental extinction spectra of the dimer arrays with different interparticle distance $d$ operating at the first- and second-order resonances are displayed in the Figs.\,\ref{fig1}e and \ref{fig1}f, respectively. In a good agreement with the modeling predictions,  the resonance is observed around 650\,nm for the 100-nm-long rods operating at the first-order SPR. For the 400-nm-long rods, which operate at the second-order SPR, the resonance is also found in the same spectral range. The electric field $z$-projection  distribution  is represented in Figs.\,\ref{fig1}(g) and \ref{fig1}(h) indicating the two- and four-antinode structure corresponding to the first- and second-order SPRs, respectively.

\begin{figure}
\includegraphics[width=\columnwidth]{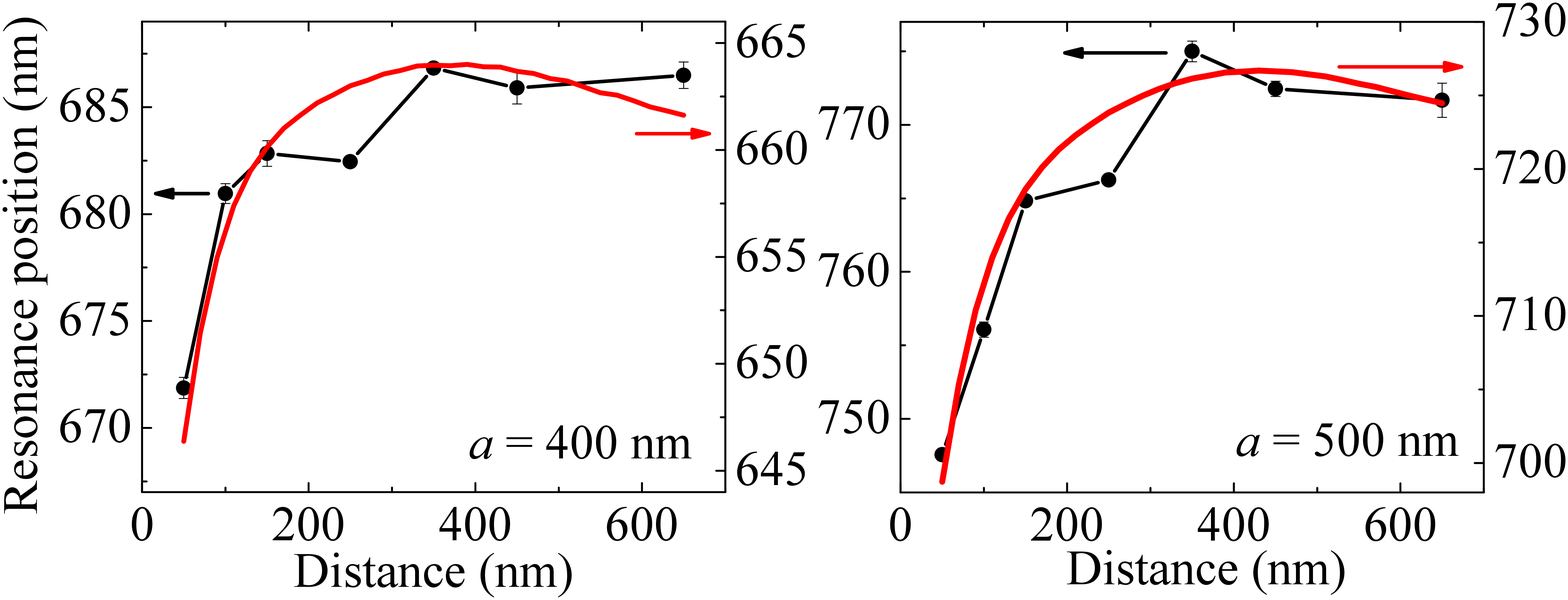}
\caption{Measured (connected dots) and calculated (solid curves) position $\lambda_0$ of the {\it second-order} longitudinal dipolar plasmon resonance of nanorod dimers as a function of the interparticle distance $d$  for  the sets of 400\,nm- and 500\,nm-long nanorods.}
\label{fig2}
\end{figure}

The extinction spectra displayed in Figs.\,\ref{fig1}(e-f) show that the resonance wavelength $\lambda_0$ depends on the interparticle separation distance $d$. This is attributed to the  effect of optical coupling between the nanorods. The experimentally observed and numerically calculated variation of the resonance wavelength $\lambda_0$ as a function of $d$ for 400-nm- and 500-nm-long nanorods operating at the second-order SPR is shown in Fig.\,\ref{fig2}. Aside the 10\% difference in the absolute value of resonance wavelength, experimental and modeling results are in good overall agreement. The difference in the absolute resonance wavelength values is due to the substrate that was not taken into account in the calculations.

Sensitivity of $\lambda_0$ to $d$ was previously shown to monotonically  depend on $d$ \cite{Weber,Dubrovina}. $\lambda_0$ is more sensitive to the variation of $d$ if the latter is small, and less sensitive for larger $d$ values. A closer inspection of Fig.\,\ref{fig2} reveals that there are some small but persistent deviations between the experimental and modeled results. While the dependence is monotonic in the modeled case, it shows some oscillatory behavior in the experiment. The oscillatory-type features in the experimental dependences are the  consequence of periodical arrangement of the dimers in a regular 2D array that leads to diffraction coupling between the nanorods. A similar effect related to the interference between the SPR and diffraction order for a 2D array of nanoantennas was reported in Refs.[\onlinecite{Teperik},\onlinecite{Lamprecht}]. This effect is discussed in more details in the case of the plasmon rule operating in a first order where its influence is much more dramatic.

\begin{figure}
\includegraphics[width=\columnwidth]{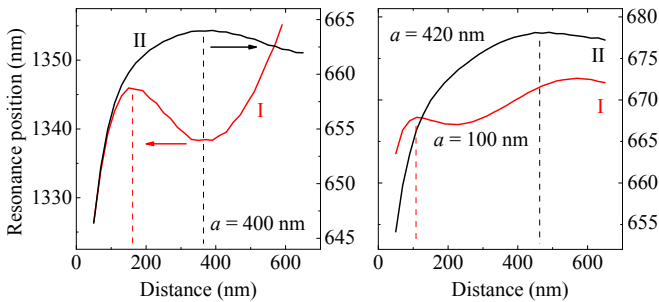}
\caption{Calculated position of the first-order (I) and second-order (II) longitudinal dipolar plasmon resonances of nanorod dimers as a function of the interparticle distance $d$. (Left): for the nanorods of the same length of $a=400$\,nm. (Right): for the nanorods of different lengths of $a=420$\,nm and $a=100$\,nm. Note the right panel having the same scale for both curves. The dashed lines show the  upper limit of the plasmon ruler working range based on the particular dimers.}
\label{fig3}
\end{figure}

The working range of a plasmon ruler, that is, the range from $d=0$ to the value where the first derivative of the $\lambda_0(d)$ function becomes equal to zero, is the key parameter showing the maximum distance that can be measured with such a ruler. To point out the key difference in $\lambda_0(d)$ for the I and II resonances, calculations of $\lambda_0(d)$ were carried out for the dimer with $a=400$\,nm for both resonance orders as seen in the left panel of Fig.\,\ref{fig3}. The working range of the first-order-resonance rulers is limited to the value of approximately 170\,nm. For $d>170$\,nm the inter-particle far-field effects are believed to disrupt the monotonous dependence leading to oscillatory behavior. On the other hand, for the second-order resonance the monotonous decay of the $\lambda_0(d)$ function is extended up to 370\,nm. This makes it more beneficial to measure large distance and length values with the second-order resonance of such a ruler. One could argue that the characteristic decay function of $\lambda_0(d)$ scales with the wavelength, and, therefore, it is not feasible to compare graphs in the left panel of Fig.\,\ref{fig3}. However, the same rule applies if the two resonances are considered on the same wavelength scale as shown in the right panel of Fig.\,\ref{fig3}. Here, the $\lambda_0(d)$ curves are given for the length values of $a=420$\,nm (second-order resonance) and $a=100$\,nm (first-order resonance). This graph straightforwardly indicates the difference between the first- and second-order resonances in plasmonic nanorod dimers. High sensitivity of the first-order resonance of a nanoparticle in the presence of the other one is understood---the resonance is ``brighter'' than the second-order one due to larger electric dipole value; the overall absolute extinction coefficient of light is approximately 3--4 times larger for the first-order resonance than for the second-order one. The same feature of the far-field coupling of dark modes was found in Ref.[\onlinecite{Teperik}]: darker (quadrupolar) modes of plasmonic cavities are less subject to far-field coupling that the brighter (dipolar) ones. 

\begin{figure}
\includegraphics[width=\columnwidth]{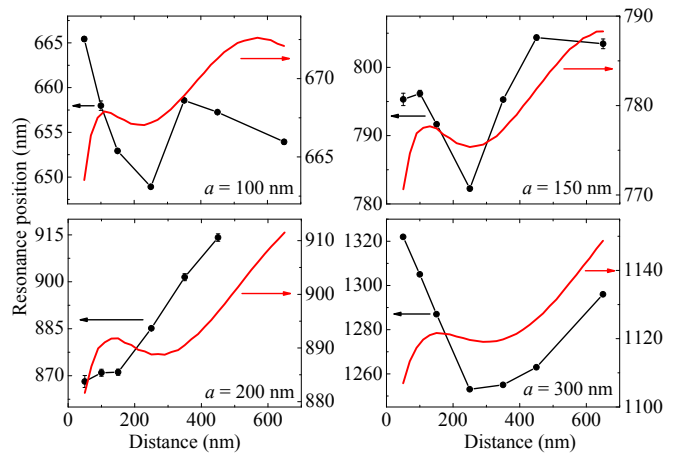}
\caption{Measured (connected dots) and calculated (solid curves) position of the {\it first-order} longitudinal dipolar plasmon resonance $\lambda_0$ of nanorod dimers as a function of the interparticle distance $d$ for the sets of 100\,nm-, 150\,nm-, 200\,nm-, and 300\,nm-long nanorods.}
\label{fig4}
\end{figure}

\begin{figure}
\includegraphics[width=0.95\columnwidth]{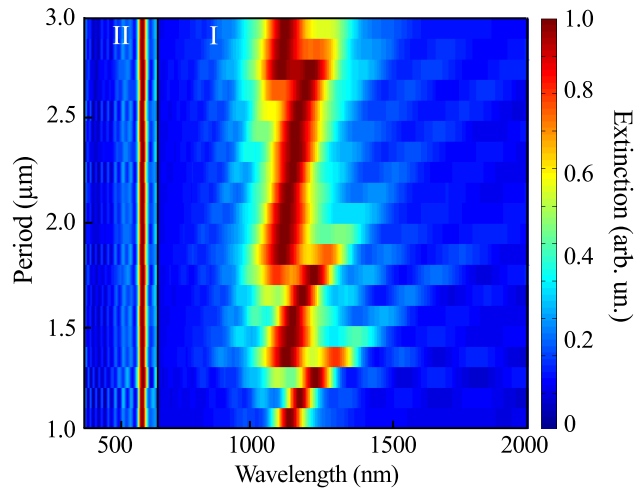}
\caption{Normalized calculated extinction spectra of the dimers with $a=d=300$\,nm as a function of the packing period. The first-order and the second-order longitudinal dipolar plasmon resonances are indicated with I and II, respectively. For the sake of clarity, the spectra are normalized separately; the border between the two normalization regions at $\lambda=650$\,nm is denoted with a black vertical  line.}
\label{fig5}
\end{figure}

Another downside of the first-order resonance is its dependence on diffraction originating from the periodic arrangement of the dimers. The effect of unwanted diffraction coupling of the first-order resonance is demonstrated with the data provided in Fig.\,\ref{fig4} for the samples with $a=100$\,nm, 150\,nm, 200\,nm, and 300\,nm. It is seen that for all the samples the shape of the experimental $\lambda_0(d)$ dependence does not reproduce the one obtained in calculations with PML boundary conditions. In order to illustrate the difference of I and II resonances in terms of diffraction coupling the effect of the packing period on extinction is shown in Fig.\,\ref{fig5}. Here, extinction of the nanorod dimer arrays with $a=300$\,nm and $d=300$\,nm is plotted as a function of wavelength and arrangement period as calculated with FDTD using periodic boundary conditions. It is seen that for the first-order resonance located in the infrared range there is a strong influence of diffraction coupling between the rods. The affection of diffraction on the resonant wavelength of plasmons was reported previously for the first-order resonance \cite{Lamprecht}. As stated above, this effect is not manifested in such a strong fashion for the second-order resonance. One can see that the second-order resonance situated at approximately 600\,nm is almost unaffected by the period alteration, and its position is preserved within the uncertainty of about 5\,nm. In contrast, the first order resonance is seen to be considerably shifted within the range from 1100 to 1300\,nm. The feature of the second-order resonance is advantageous in terms of general stability of the plasmon ruler to diffraction coupling, which could be critical in plasmon rulers allocated in periodical arrangements.

%
%
%\section{discussion}
%
%Points for discussion: why 2nd res 
%
%
%In this section we discuss the results from the point of view of comparing the $\lambda(d)$ dependences for the first- and second-order resonances. It would be good if we could supply some analytical dependences here.
%
%I think we should insert the graph with the period variation here. It tell a lot about ``far-field stability'' of the SPR (fig16 from the course work).

%
%
%\section{conclusion + acknowledgements}
In conclusion, the use of the second-order plasmon resonance in gold nanorod $\pi$-dimers for plasmon rulers is justified experimentally and theoretically.  Explicit evidences are provided to demonstrate the essential difference between $\pi$-dimer nanorod rulers operating at the first- and second-order surface plasmon resonances. The maximum distance that can be measured using a ruler operating at the second-order resonance is extended up to 400\,nm as compared to about 100\,nm provided by the ruler based on the first-order resonance. Moreover, the second-order-resonance ruler is shown to merely suffer from diffraction coupling. This fact makes second-order surface plasmon resonances attractive for periodic-arrangement solutions involving plasmon rulers.

%This is further confirmed by comparing data for nanorods of different length. 

The authors are grateful to the IEF MINERVE clean-room staff for their help with the samples technological realization and to Denis Presnov for providing SEM micrographs of the samples. This work was supported by Russian Foundation for Basic Research and Ministry of Education and Science of Russian Federation. 

%\bibliography{APL2013}

\begin{thebibliography}{21}%
\makeatletter
\providecommand \@ifxundefined [1]{%
 \@ifx{#1\undefined}
}%
\providecommand \@ifnum [1]{%
 \ifnum #1\expandafter \@firstoftwo
 \else \expandafter \@secondoftwo
 \fi
}%
\providecommand \@ifx [1]{%
 \ifx #1\expandafter \@firstoftwo
 \else \expandafter \@secondoftwo
 \fi
}%
\providecommand \natexlab [1]{#1}%
\providecommand \enquote  [1]{``#1''}%
\providecommand \bibnamefont  [1]{#1}%
\providecommand \bibfnamefont [1]{#1}%
\providecommand \citenamefont [1]{#1}%
\providecommand \href@noop [0]{\@secondoftwo}%
\providecommand \href [0]{\begingroup \@sanitize@url \@href}%
\providecommand \@href[1]{\@@startlink{#1}\@@href}%
\providecommand \@@href[1]{\endgroup#1\@@endlink}%
\providecommand \@sanitize@url [0]{\catcode `\\12\catcode `\$12\catcode
  `\&12\catcode `\#12\catcode `\^12\catcode `\_12\catcode `\%12\relax}%
\providecommand \@@startlink[1]{}%
\providecommand \@@endlink[0]{}%
\providecommand \url  [0]{\begingroup\@sanitize@url \@url }%
\providecommand \@url [1]{\endgroup\@href {#1}{\urlprefix }}%
\providecommand \urlprefix  [0]{URL }%
\providecommand \Eprint [0]{\href }%
\providecommand \doibase [0]{http://dx.doi.org/}%
\providecommand \selectlanguage [0]{\@gobble}%
\providecommand \bibinfo  [0]{\@secondoftwo}%
\providecommand \bibfield  [0]{\@secondoftwo}%
\providecommand \translation [1]{[#1]}%
\providecommand \BibitemOpen [0]{}%
\providecommand \bibitemStop [0]{}%
\providecommand \bibitemNoStop [0]{.\EOS\space}%
\providecommand \EOS [0]{\spacefactor3000\relax}%
\providecommand \BibitemShut  [1]{\csname bibitem#1\endcsname}%
\let\auto@bib@innerbib\@empty
%</preamble>
\bibitem [{\citenamefont {Brown}\ \emph {et~al.}(2010)\citenamefont {Brown},
  \citenamefont {Nativo}, \citenamefont {Smith}, \citenamefont {Stirling},
  \citenamefont {Edwards}, \citenamefont {Venugopal}, \citenamefont {Flint},
  \citenamefont {Plumb}, \citenamefont {Graham},\ and\ \citenamefont
  {Wheate}}]{Brown}%
  \BibitemOpen
  \bibfield  {author} {\bibinfo {author} {\bibfnamefont {S.~D.}\ \bibnamefont
  {Brown}}, \bibinfo {author} {\bibfnamefont {P.}~\bibnamefont {Nativo}},
  \bibinfo {author} {\bibfnamefont {J.-A.}\ \bibnamefont {Smith}}, \bibinfo
  {author} {\bibfnamefont {D.}~\bibnamefont {Stirling}}, \bibinfo {author}
  {\bibfnamefont {P.~R.}\ \bibnamefont {Edwards}}, \bibinfo {author}
  {\bibfnamefont {B.}~\bibnamefont {Venugopal}}, \bibinfo {author}
  {\bibfnamefont {D.~J.}\ \bibnamefont {Flint}}, \bibinfo {author}
  {\bibfnamefont {J.~A.}\ \bibnamefont {Plumb}}, \bibinfo {author}
  {\bibfnamefont {D.}~\bibnamefont {Graham}}, \ and\ \bibinfo {author}
  {\bibfnamefont {N.~J.}\ \bibnamefont {Wheate}},\ }\href@noop {} {\bibfield
  {journal} {\bibinfo  {journal} {Journal of the American Chemical Society}\
  }\textbf {\bibinfo {volume} {132}},\ \bibinfo {pages} {4678--4684} (\bibinfo
  {year} {2010})}\BibitemShut {NoStop}%
\bibitem [{\citenamefont {Perrault}\ and\ \citenamefont
  {Chan}(2010)}]{Perrault}%
  \BibitemOpen
  \bibfield  {author} {\bibinfo {author} {\bibfnamefont {S.~D.}\ \bibnamefont
  {Perrault}}\ and\ \bibinfo {author} {\bibfnamefont {W.~C.~W.}\ \bibnamefont
  {Chan}},\ }\href@noop {} {\bibfield  {journal} {\bibinfo  {journal}
  {Proceedings of the National Academy of Sciences}\ }\textbf {\bibinfo
  {volume} {107}},\ \bibinfo {pages} {11194--11199} (\bibinfo {year}
  {2010})}\BibitemShut {NoStop}%
\bibitem [{\citenamefont {Huang}\ \emph {et~al.}(2006)\citenamefont {Huang},
  \citenamefont {El-Sayed}, \citenamefont {Qian},\ and\ \citenamefont
  {El-Sayed}}]{Huang}%
  \BibitemOpen
  \bibfield  {author} {\bibinfo {author} {\bibfnamefont {X.}~\bibnamefont
  {Huang}}, \bibinfo {author} {\bibfnamefont {I.~H.}\ \bibnamefont {El-Sayed}},
  \bibinfo {author} {\bibfnamefont {W.}~\bibnamefont {Qian}}, \ and\ \bibinfo
  {author} {\bibfnamefont {M.~A.}\ \bibnamefont {El-Sayed}},\ }\href@noop {}
  {\bibfield  {journal} {\bibinfo  {journal} {Journal of the American Chemical
  Society}\ }\textbf {\bibinfo {volume} {128}},\ \bibinfo {pages} {2115--2120}
  (\bibinfo {year} {2006})}\BibitemShut {NoStop}%
\bibitem [{\citenamefont {Reinhard}\ \emph {et~al.}(2005)\citenamefont
  {Reinhard}, \citenamefont {Siu}, \citenamefont {Agarwal}, \citenamefont
  {Alivisatos},\ and\ \citenamefont {Liphardt}}]{Reinhard}%
  \BibitemOpen
  \bibfield  {author} {\bibinfo {author} {\bibfnamefont {B.~M.}\ \bibnamefont
  {Reinhard}}, \bibinfo {author} {\bibfnamefont {M.}~\bibnamefont {Siu}},
  \bibinfo {author} {\bibfnamefont {H.}~\bibnamefont {Agarwal}}, \bibinfo
  {author} {\bibfnamefont {A.~P.}\ \bibnamefont {Alivisatos}}, \ and\ \bibinfo
  {author} {\bibfnamefont {J.}~\bibnamefont {Liphardt}},\ }\href@noop {}
  {\bibfield  {journal} {\bibinfo  {journal} {Nano Letters}\ }\textbf {\bibinfo
  {volume} {5}},\ \bibinfo {pages} {2246--2252} (\bibinfo {year}
  {2005})}\BibitemShut {NoStop}%
\bibitem [{\citenamefont {Jain}, \citenamefont {Huang},\ and\ \citenamefont
  {El-Sayed}(2007)}]{JainEquation}%
  \BibitemOpen
  \bibfield  {author} {\bibinfo {author} {\bibfnamefont {P.~K.}\ \bibnamefont
  {Jain}}, \bibinfo {author} {\bibfnamefont {W.}~\bibnamefont {Huang}}, \ and\
  \bibinfo {author} {\bibfnamefont {M.~A.}\ \bibnamefont {El-Sayed}},\
  }\href@noop {} {\bibfield  {journal} {\bibinfo  {journal} {Nano Letters}\
  }\textbf {\bibinfo {volume} {7}},\ \bibinfo {pages} {2080--2088} (\bibinfo
  {year} {2007})}\BibitemShut {NoStop}%
\bibitem [{\citenamefont {Rechberger}\ \emph {et~al.}(2003)\citenamefont
  {Rechberger}, \citenamefont {Hohenau}, \citenamefont {Leitner}, \citenamefont
  {Krenn}, \citenamefont {Lamprecht},\ and\ \citenamefont
  {Aussenegg}}]{Rechberger}%
  \BibitemOpen
  \bibfield  {author} {\bibinfo {author} {\bibfnamefont {W.}~\bibnamefont
  {Rechberger}}, \bibinfo {author} {\bibfnamefont {A.}~\bibnamefont {Hohenau}},
  \bibinfo {author} {\bibfnamefont {A.}~\bibnamefont {Leitner}}, \bibinfo
  {author} {\bibfnamefont {J.~R.}\ \bibnamefont {Krenn}}, \bibinfo {author}
  {\bibfnamefont {B.}~\bibnamefont {Lamprecht}}, \ and\ \bibinfo {author}
  {\bibfnamefont {F.~R.}\ \bibnamefont {Aussenegg}},\ }\href@noop {} {\bibfield
   {journal} {\bibinfo  {journal} {Optics Communications}\ }\textbf {\bibinfo
  {volume} {220}},\ \bibinfo {pages} {137--141} (\bibinfo {year}
  {2003})}\BibitemShut {NoStop}%
\bibitem [{\citenamefont {Tsai}\ \emph {et~al.}(2012)\citenamefont {Tsai},
  \citenamefont {Lin}, \citenamefont {Wu}, \citenamefont {Lin}, \citenamefont
  {Lu},\ and\ \citenamefont {Lee}}]{ChiayangTsai}%
  \BibitemOpen
  \bibfield  {author} {\bibinfo {author} {\bibfnamefont {C.-Y.}\ \bibnamefont
  {Tsai}}, \bibinfo {author} {\bibfnamefont {J.-W.}\ \bibnamefont {Lin}},
  \bibinfo {author} {\bibfnamefont {C.-Y.}\ \bibnamefont {Wu}}, \bibinfo
  {author} {\bibfnamefont {P.-T.}\ \bibnamefont {Lin}}, \bibinfo {author}
  {\bibfnamefont {T.-W.}\ \bibnamefont {Lu}}, \ and\ \bibinfo {author}
  {\bibfnamefont {P.-T.}\ \bibnamefont {Lee}},\ }\href@noop {} {\bibfield
  {journal} {\bibinfo  {journal} {Nano Letters}\ }\textbf {\bibinfo {volume}
  {12}},\ \bibinfo {pages} {1648--1654} (\bibinfo {year} {2012})}\BibitemShut
  {NoStop}%
\bibitem [{\citenamefont {Tabor}\ \emph {et~al.}(2009)\citenamefont {Tabor},
  \citenamefont {Murali}, \citenamefont {Mahmoud},\ and\ \citenamefont
  {El-Sayed}}]{Tabor}%
  \BibitemOpen
  \bibfield  {author} {\bibinfo {author} {\bibfnamefont {C.}~\bibnamefont
  {Tabor}}, \bibinfo {author} {\bibfnamefont {R.}~\bibnamefont {Murali}},
  \bibinfo {author} {\bibfnamefont {M.}~\bibnamefont {Mahmoud}}, \ and\
  \bibinfo {author} {\bibfnamefont {M.~A.}\ \bibnamefont {El-Sayed}},\
  }\href@noop {} {\bibfield  {journal} {\bibinfo  {journal} {The Journal of
  Physical Chemistry A}\ }\textbf {\bibinfo {volume} {113}},\ \bibinfo {pages}
  {1946--1953} (\bibinfo {year} {2009})}\BibitemShut {NoStop}%
\bibitem [{\citenamefont {Yang}\ \emph {et~al.}(2010)\citenamefont {Yang},
  \citenamefont {Kobori}, \citenamefont {He}, \citenamefont {Lin},
  \citenamefont {Chen}, \citenamefont {Li}, \citenamefont {Kanehara},
  \citenamefont {Teranishi},\ and\ \citenamefont {Gwo}}]{ShuchunYang}%
  \BibitemOpen
  \bibfield  {author} {\bibinfo {author} {\bibfnamefont {S.-C.}\ \bibnamefont
  {Yang}}, \bibinfo {author} {\bibfnamefont {H.}~\bibnamefont {Kobori}},
  \bibinfo {author} {\bibfnamefont {C.-L.}\ \bibnamefont {He}}, \bibinfo
  {author} {\bibfnamefont {M.-H.}\ \bibnamefont {Lin}}, \bibinfo {author}
  {\bibfnamefont {H.-Y.}\ \bibnamefont {Chen}}, \bibinfo {author}
  {\bibfnamefont {C.}~\bibnamefont {Li}}, \bibinfo {author} {\bibfnamefont
  {M.}~\bibnamefont {Kanehara}}, \bibinfo {author} {\bibfnamefont
  {T.}~\bibnamefont {Teranishi}}, \ and\ \bibinfo {author} {\bibfnamefont
  {S.}~\bibnamefont {Gwo}},\ }\href@noop {} {\bibfield  {journal} {\bibinfo
  {journal} {Nano Letters}\ }\textbf {\bibinfo {volume} {10}},\ \bibinfo
  {pages} {632--637} (\bibinfo {year} {2010})}\BibitemShut {NoStop}%
\bibitem [{\citenamefont {Funston}\ \emph {et~al.}(2009)\citenamefont
  {Funston}, \citenamefont {Novo}, \citenamefont {Davis},\ and\ \citenamefont
  {Mulvaney}}]{Funston}%
  \BibitemOpen
  \bibfield  {author} {\bibinfo {author} {\bibfnamefont {A.~M.}\ \bibnamefont
  {Funston}}, \bibinfo {author} {\bibfnamefont {C.}~\bibnamefont {Novo}},
  \bibinfo {author} {\bibfnamefont {T.~J.}\ \bibnamefont {Davis}}, \ and\
  \bibinfo {author} {\bibfnamefont {P.}~\bibnamefont {Mulvaney}},\ }\href@noop
  {} {\bibfield  {journal} {\bibinfo  {journal} {Nano Letters}\ }\textbf
  {\bibinfo {volume} {9}},\ \bibinfo {pages} {1651--1658} (\bibinfo {year}
  {2009})}\BibitemShut {NoStop}%
\bibitem [{\citenamefont {Liu}\ \emph {et~al.}(2008)\citenamefont {Liu},
  \citenamefont {Boltasseva}, \citenamefont {Pedersen}, \citenamefont {Bakker},
  \citenamefont {Kildishev}, \citenamefont {Drachev},\ and\ \citenamefont
  {Shalaev}}]{ZhengtongLiu}%
  \BibitemOpen
  \bibfield  {author} {\bibinfo {author} {\bibfnamefont {Z.}~\bibnamefont
  {Liu}}, \bibinfo {author} {\bibfnamefont {A.}~\bibnamefont {Boltasseva}},
  \bibinfo {author} {\bibfnamefont {R.~H.}\ \bibnamefont {Pedersen}}, \bibinfo
  {author} {\bibfnamefont {R.}~\bibnamefont {Bakker}}, \bibinfo {author}
  {\bibfnamefont {A.~V.}\ \bibnamefont {Kildishev}}, \bibinfo {author}
  {\bibfnamefont {V.~P.}\ \bibnamefont {Drachev}}, \ and\ \bibinfo {author}
  {\bibfnamefont {V.~M.}\ \bibnamefont {Shalaev}},\ }\href@noop {} {\bibfield
  {journal} {\bibinfo  {journal} {Metamaterials}\ }\textbf {\bibinfo {volume}
  {2}},\ \bibinfo {pages} {45--51} (\bibinfo {year} {2008})}\BibitemShut
  {NoStop}%
\bibitem [{\citenamefont {Alexander}\ \emph {et~al.}(2010)\citenamefont
  {Alexander}, \citenamefont {Skinner}, \citenamefont {Zhang}, \citenamefont
  {Wei},\ and\ \citenamefont {Lopez}}]{Alexander}%
  \BibitemOpen
  \bibfield  {author} {\bibinfo {author} {\bibfnamefont {K.~D.}\ \bibnamefont
  {Alexander}}, \bibinfo {author} {\bibfnamefont {K.}~\bibnamefont {Skinner}},
  \bibinfo {author} {\bibfnamefont {S.}~\bibnamefont {Zhang}}, \bibinfo
  {author} {\bibfnamefont {H.}~\bibnamefont {Wei}}, \ and\ \bibinfo {author}
  {\bibfnamefont {R.}~\bibnamefont {Lopez}},\ }\href@noop {} {\bibfield
  {journal} {\bibinfo  {journal} {Nano Letters}\ }\textbf {\bibinfo {volume}
  {10}},\ \bibinfo {pages} {4488--4493} (\bibinfo {year} {2010})}\BibitemShut
  {NoStop}%
\bibitem [{\citenamefont {Podolsky}, \citenamefont {Sarychev},\ and\
  \citenamefont {Shalaev}(2002)}]{Podolsky}%
  \BibitemOpen
  \bibfield  {author} {\bibinfo {author} {\bibfnamefont {V.~A.}\ \bibnamefont
  {Podolsky}}, \bibinfo {author} {\bibfnamefont {A.~K.}\ \bibnamefont
  {Sarychev}}, \ and\ \bibinfo {author} {\bibfnamefont {V.~M.}\ \bibnamefont
  {Shalaev}},\ }\href@noop {} {\bibfield  {journal} {\bibinfo  {journal}
  {Journal of Nonlinear Optical Physics \& Materials}\ }\textbf {\bibinfo
  {volume} {11}},\ \bibinfo {pages} {65--74} (\bibinfo {year}
  {2002})}\BibitemShut {NoStop}%
\bibitem [{\citenamefont {Payne}\ \emph {et~al.}(2006)\citenamefont {Payne},
  \citenamefont {Shuford}, \citenamefont {Park}, \citenamefont {Schatz},\ and\
  \citenamefont {Mirkin}}]{Payne}%
  \BibitemOpen
  \bibfield  {author} {\bibinfo {author} {\bibfnamefont {E.~K.}\ \bibnamefont
  {Payne}}, \bibinfo {author} {\bibfnamefont {K.~L.}\ \bibnamefont {Shuford}},
  \bibinfo {author} {\bibfnamefont {S.}~\bibnamefont {Park}}, \bibinfo {author}
  {\bibfnamefont {G.~C.}\ \bibnamefont {Schatz}}, \ and\ \bibinfo {author}
  {\bibfnamefont {C.~A.}\ \bibnamefont {Mirkin}},\ }\href@noop {} {\bibfield
  {journal} {\bibinfo  {journal} {The Journal of Physical Chemistry B}\
  }\textbf {\bibinfo {volume} {110}},\ \bibinfo {pages} {2150--2154} (\bibinfo
  {year} {2006})}\BibitemShut {NoStop}%
\bibitem [{\citenamefont {Jain}, \citenamefont {Eustis},\ and\ \citenamefont
  {El-Sayed}(2006)}]{JainNR}%
  \BibitemOpen
  \bibfield  {author} {\bibinfo {author} {\bibfnamefont {P.~K.}\ \bibnamefont
  {Jain}}, \bibinfo {author} {\bibfnamefont {S.}~\bibnamefont {Eustis}}, \ and\
  \bibinfo {author} {\bibfnamefont {M.~A.}\ \bibnamefont {El-Sayed}},\
  }\href@noop {} {\bibfield  {journal} {\bibinfo  {journal} {Journal of
  Physical Chemistry B}\ }\textbf {\bibinfo {volume} {110}},\ \bibinfo {pages}
  {18243} (\bibinfo {year} {2006})}\BibitemShut {NoStop}%
\bibitem [{\citenamefont {Willingham}, \citenamefont {Brandl},\ and\
  \citenamefont {Nordlander}(2008)}]{Willingham}%
  \BibitemOpen
  \bibfield  {author} {\bibinfo {author} {\bibfnamefont {B.}~\bibnamefont
  {Willingham}}, \bibinfo {author} {\bibfnamefont {D.~W.}\ \bibnamefont
  {Brandl}}, \ and\ \bibinfo {author} {\bibfnamefont {P.}~\bibnamefont
  {Nordlander}},\ }\href@noop {} {\bibfield  {journal} {\bibinfo  {journal}
  {Applied Physics B}\ }\textbf {\bibinfo {volume} {93}},\ \bibinfo {pages}
  {209--216} (\bibinfo {year} {2008})}\BibitemShut {NoStop}%
\bibitem [{\citenamefont {Palik}(1985)}]{Palik}%
  \BibitemOpen
  \bibfield  {author} {\bibinfo {author} {\bibfnamefont {E.~D.}\ \bibnamefont
  {Palik}},\ }\href@noop {} {\emph {\bibinfo {title} {Handbook of Optical
  Constants of Solids}}}\ (\bibinfo  {publisher} {Academic Press, Boston},\
  \bibinfo {year} {1985})\BibitemShut {NoStop}%
\bibitem [{\citenamefont {Weber}\ \emph {et~al.}(2011)\citenamefont {Weber},
  \citenamefont {Albella}, \citenamefont {Alonso-Gonzalez}, \citenamefont
  {Neubrech}, \citenamefont {Gui}, \citenamefont {Nagao}, \citenamefont
  {Hillenbrand}, \citenamefont {Aizpurua},\ and\ \citenamefont
  {Pucci}}]{Weber}%
  \BibitemOpen
  \bibfield  {author} {\bibinfo {author} {\bibfnamefont {D.}~\bibnamefont
  {Weber}}, \bibinfo {author} {\bibfnamefont {P.}~\bibnamefont {Albella}},
  \bibinfo {author} {\bibfnamefont {P.}~\bibnamefont {Alonso-Gonzalez}},
  \bibinfo {author} {\bibfnamefont {F.}~\bibnamefont {Neubrech}}, \bibinfo
  {author} {\bibfnamefont {H.}~\bibnamefont {Gui}}, \bibinfo {author}
  {\bibfnamefont {T.}~\bibnamefont {Nagao}}, \bibinfo {author} {\bibfnamefont
  {R.}~\bibnamefont {Hillenbrand}}, \bibinfo {author} {\bibfnamefont
  {J.}~\bibnamefont {Aizpurua}}, \ and\ \bibinfo {author} {\bibfnamefont
  {A.}~\bibnamefont {Pucci}},\ }\href@noop {} {\bibfield  {journal} {\bibinfo
  {journal} {Optics Express}\ }\textbf {\bibinfo {volume} {19}},\ \bibinfo
  {pages} {15047--15061} (\bibinfo {year} {2011})}\BibitemShut {NoStop}%
\bibitem [{\citenamefont {Dubrovina}\ \emph {et~al.}(2012)\citenamefont
  {Dubrovina}, \citenamefont {Cunff}, \citenamefont {Burokur}, \citenamefont
  {Ghasemi}, \citenamefont {Degiron}, \citenamefont {Lustrac}, \citenamefont
  {Vial}, \citenamefont {Lerondel},\ and\ \citenamefont {Lupu}}]{Dubrovina}%
  \BibitemOpen
  \bibfield  {author} {\bibinfo {author} {\bibfnamefont {N.}~\bibnamefont
  {Dubrovina}}, \bibinfo {author} {\bibfnamefont {L.~O.}\ \bibnamefont
  {Cunff}}, \bibinfo {author} {\bibfnamefont {N.}~\bibnamefont {Burokur}},
  \bibinfo {author} {\bibfnamefont {R.}~\bibnamefont {Ghasemi}}, \bibinfo
  {author} {\bibfnamefont {A.}~\bibnamefont {Degiron}}, \bibinfo {author}
  {\bibfnamefont {A.}~\bibnamefont {Lustrac}}, \bibinfo {author} {\bibfnamefont
  {A.}~\bibnamefont {Vial}}, \bibinfo {author} {\bibfnamefont {G.}~\bibnamefont
  {Lerondel}}, \ and\ \bibinfo {author} {\bibfnamefont {A.}~\bibnamefont
  {Lupu}},\ }\href@noop {} {\bibfield  {journal} {\bibinfo  {journal} {Applied
  Physics A}\ }\textbf {\bibinfo {volume} {109}},\ \bibinfo {pages} {901--906}
  (\bibinfo {year} {2012})}\BibitemShut {NoStop}%
\bibitem [{\citenamefont {Teperik}\ and\ \citenamefont
  {Degiron}(2012)}]{Teperik}%
  \BibitemOpen
  \bibfield  {author} {\bibinfo {author} {\bibfnamefont {T.~V.}\ \bibnamefont
  {Teperik}}\ and\ \bibinfo {author} {\bibfnamefont {A.}~\bibnamefont
  {Degiron}},\ }\href@noop {} {\bibfield  {journal} {\bibinfo  {journal}
  {Physical Review Letters}\ }\textbf {\bibinfo {volume} {108}},\ \bibinfo
  {pages} {147401} (\bibinfo {year} {2012})}\BibitemShut {NoStop}%
\bibitem [{\citenamefont {Lamprecht}\ \emph {et~al.}(2000)\citenamefont
  {Lamprecht}, \citenamefont {Schider}, \citenamefont {Lechner}, \citenamefont
  {Ditlbacher}, \citenamefont {Krenn}, \citenamefont {Leitner},\ and\
  \citenamefont {Aussenegg}}]{Lamprecht}%
  \BibitemOpen
  \bibfield  {author} {\bibinfo {author} {\bibfnamefont {B.}~\bibnamefont
  {Lamprecht}}, \bibinfo {author} {\bibfnamefont {G.}~\bibnamefont {Schider}},
  \bibinfo {author} {\bibfnamefont {R.~T.}\ \bibnamefont {Lechner}}, \bibinfo
  {author} {\bibfnamefont {H.}~\bibnamefont {Ditlbacher}}, \bibinfo {author}
  {\bibfnamefont {J.~R.}\ \bibnamefont {Krenn}}, \bibinfo {author}
  {\bibfnamefont {A.}~\bibnamefont {Leitner}}, \ and\ \bibinfo {author}
  {\bibfnamefont {F.~R.}\ \bibnamefont {Aussenegg}},\ }\href@noop {} {\bibfield
   {journal} {\bibinfo  {journal} {Physical Review Letters}\ }\textbf {\bibinfo
  {volume} {84}},\ \bibinfo {pages} {4721--4724} (\bibinfo {year}
  {2000})}\BibitemShut {NoStop}%
\end{thebibliography}
%\bibliographystyle{dokdok}


\end{document}